\documentclass[aps,prb,twocolumn,groupedaddress,showpacs]{revtex4}
\usepackage{graphicx}
\newcommand{\be}{\begin{equation}}
\newcommand{\ee}{\end{equation}}
\newcommand{\bea}{\begin{eqnarray}}
\newcommand{\eea}{\end{eqnarray}}

\begin{document}

\title{ Scaling behavior of the energy gap of spin-$\frac{1}{2}$ AF-Heisenberg chain in both uniform and staggered fields }

\author{ S. Mahdavifar }
\address{Institute for Advanced Studies in Basic Sciences, Zanjan 45195-1159,
Iran}
\email{mahdavifar@iasbs.ac.ir}

\begin{abstract}

  We have studied the energy gap of the 1D AF-Heisenberg model in the
  presence of both uniform ($H$) and staggered ($h$) magnetic fields using the exact
  diagonalization technique. We have found that the opening of the gap in the presence of a staggered field scales with $h^{\nu}$, where $\nu=\nu(H)$ is the critical exponent and depends on the uniform field. With respect to the range of the staggered magnetic field, we have identified two
  regimes through which the $H$-dependence of the real
  critical exponent $\nu(H)$ can be numerically calculated. Our numerical
  results are in good agreement with the results obtained by
  theoretical approaches.

\end{abstract}

\maketitle

\section{Introduction}

   The effect of external magnetic fields in the quantum properties of 
   low-dimensional magnets has been of much interest
   in recent years. Experimental and theoretical studies of these systems have revealed a plethora of quantum flactuation 
   phenomena, not usally observed in higher dimensions. The magnetization processes in antiferromagnetic (AF) 
   spin chains and ladders have been under intensive investigation using novel numerical techniques. The
   progress in the experimental front is achived by introduction of high-field neutron scattering studies and 
   synthesis of magnetic quasi-one dimensional systems such as the spin-$\frac{1}{2}$ antiferromagnet
   Cu benzoate \cite{dender, asano1, asano2} and $Yb_{4}As_{3}$ \cite{oshikawa1, shiba, kohgi}. Due to 
   these developments we can now observe the effect of a staggered magnetic field (or even more complicated interactions) on the low energy behavior of a one-dimentional quantum model in the laboratory.

   There exist different mechanisms for generating a staggered field in a real magnet \cite{oshikawa2, wang, sato}. In Cu benzoate the alternating crystal axes is the source of such a field. Dender et.al. \cite{dender} showed that an effective staggered field can be generated by the alternating g-tensor. Theoreticaly, Afflec et.al. \cite{oshikawa2} have studied how an effective staggered field is generated by Dzyaloshinskii-Moriya (DM) interaction if the crystal symmetry is sufficiently low. They showed that in the presence of DM interaction along the AF chain, an applied uniform field $\overrightarrow{H}$
 generates an effective staggered field $\overrightarrow{h}$. Ignoring small residual anisotropies, they obtained an effective hamiltonian where a one-dimensional Heisenberg AF chain is placed in perpendicular uniform ($H$) and staggered ($h$) fields
\begin{equation}
\hat{H}= \sum_{j} [J \overrightarrow{S}_{j}.
\overrightarrow{S}_{j+1}-H S_{j}^{x}+h (-1)^{j}
S_{j}^{z}]\label{Hamiltonian}
\end{equation}

 It is expected \cite{oshikawa2, alcaraz1} that the staggered field induces an
 excitation gap in the $S=\frac{1}{2}$ Heisenberg
 antiferromagnetic (AF) chain, which should be otherwise gapless.
 Such as excitation gap caused by the staggered field is indeed found
 in real magnets \cite{dender, kohgi, feyerherm}.

 In the absence of the staggered magnetic field ($h=0$) and the
 uniform magnetic field ($H=0$), the spectrum is gapless. In the ground state, the system is in the
 spin-fluid phase, where the decay of correlations fallow a power low. When a uniform magnetic field is 
 applied the spectrum of the system remains gapless until a critical field $H_{c}=2 J$, is reached. Here a phase transition of the
 Pokrovsky-Talapov type \cite{pokrovsky} occurs and the ground state becomes
 a completely ordered ferromagnetic state \cite{griffiths}. Since the
 uniform magnetic field does not destroy the exact integrability
 of the Heisenberg model, the eigenspectra is exactly solvable. Applying a staggered magnetic field, the integrability is lost. The application of a staggered magnetic field when $H=0$, produces an
 antiferromagnetically ordered (Neel order) ground-state and induces a
 gap in the spectrum of the model. Heisenberg model in both staggered and uniform
 fields has been recently studied \cite{lou} using density matrix renormalization group (DMRG). It is shown that bound midgap states generally exist in open boundary AF-Heisenberg chains. The gap and midgap energies in the thermodynamic limit are obtained by extrapolating numerical results of small chain sizes up to 200 sites. It is revealed that some of the gap and midgap energies for the half-integer spin chains fit well to a scaling function derived from the quantum Sine-Gordon model, but other low energy excitations do not fit equally well. 

 In this paper, we present the numerical results obtained on the low-energy
 states of the 1D AF-Heisenberg model in both uniform and staggered fields using an exact diagonalization technique for finite systems. We calculate the
 spin gap as a function of applied staggered field in the presence of small uniform field ($0\leq H<0.1$). With respect to the range of the staggered magnetic field, we show that there are two regimes in which we can compute the real critical exponent of the energy gap and it is important to note to which one of these regimes the numerical data are related. In Sec. II
 we discuss the scaling behavior of the gap using the available limiting behaviors. The leading exponent of the staggered field $h$, depends on $H$ boath in finite size and thermodynamic limit. In Sec. III, we
 explain how, in certain limits the numerical calculations may produce
 incorrect result for the critical exponent. We apply a perturbative approach\cite{langari} to find the
 correct critical exponent in the small-$x$ ($x=N h^{\nu(H)}$) regime. In Sec. IV, we
 increase the scaling parameter $x$ and find the correct critical exponent in the large-$x$ regime. Finally, the
 summary and discussion are presented in Sec. V.

\section{The Scaling Behavior of the Gap}

  In the high field neutron-scattering experiment on Cu benzoate
  \cite{dender}, which is a quasi-one dimensional $S=\frac{1}{2}$ antiferromagnet,
  the magnetic field induces a gap in excitation spectrum of the
  magnet. The observed gap is proportional to $H_{0}^{0.65}$,
  where  $H_{0}$ is the magnitude of the applied field. This exponent of about $\frac{2}{3}$ describing the field
  dependence of the gap  obtained in different experiments \cite{feyerherm, kohgi} identify the source of this gap as the staggered field.

Using  bosonization techniques, Affleck et.al showed that, the gap scales as
\begin{eqnarray}
\Delta(h, H) \sim h^{\nu(H)}, \label{e5}
\end{eqnarray}
where $\nu(H)$ is the critical exponent of the gap and when $H$ is stricly $0$, $\nu(H=0)=\frac{2}{3}$. The $H$-dependence of the exponent $\nu(H)$ is studied
numerically in Ref.[15]. Their approach is based on the
$\eta$-exponent, defined through the static structure factor
of the model in the absence of a staggered field ($h=0$). They show that there is a relation between the critical exponent of the gap and  $\eta$-exponent. Then by computing the $\eta$-exponent of the structure factor of the model, they predict the $H$-dependence of $\nu(H)$. Similarly, In an interesting recent work \cite{chernyshev}, the effect of an external field on the gap of the 2D AF Heisenberg model with DM interaction has been studied. It is shown that the effect of the external field on the gap can be predicted by investigating the on-site magnetization of the model.  

Here we study the evolution of the gap, using
the conformal estimates of the small perturbation $h\ll1$, and the
finite size scaling estimates of the energy eigenvalues of the
small chains in the presence of the staggered field ($h\neq0$). We
argue that there are two regimes in which the real critical exponent
can be numerically calculated and it is very important to note to
which one of these regimes the numerical data are related.

Let us rewrite the Hamiltonian (1) in the form
\begin{eqnarray}
\hat{H}&=&\hat{H}_{0}+V \nonumber \\
\hat{H}_{0}&=& \sum_{j} [J \overrightarrow{S}_{j}.
\overrightarrow{S}_{j+1}-H S_{j}^{x}] \nonumber \\
V&=&h\sum_{j} (-1)^{j} S_{j}^{z}, \label{efh6}
\end{eqnarray}
where $\hat{H}_{0}$ is exactly solvable by the Bethe ansatz and
the staggered field $h\ll1$ is very small. For a small
perturbation V, we can use conformal estimates. The large distance
asymptotic of the correlation function of the model in the absence
of the staggered field $(h=0)$ is obtained \cite{bogoliubovn} as
\begin{eqnarray}
\langle S_{j}^{z} S_{j+n}^{z}\rangle \sim
\frac{(-1)^{n}}{n^{\alpha(H)}}.
\end{eqnarray}
Where $\alpha(H)$ is a function of the uniform ($H$) field and is
found using the Bethe ansatz as
\begin{eqnarray}
\alpha(H) \sim 1-\frac{1}{2 \ln(1/H)},~~~~~~~~H \rightarrow 0
\end{eqnarray}
where $\alpha(0)=1$ and $\alpha(2)=\frac{1}{2}$. By investigating the perturbed
action for the model and performing an infinitesimal renormalization group with a scale
$\lambda$, one can show that the staggered magnetic field scales as $h'=h
\lambda^{2-\frac{\alpha(H)}{2}}$ which leads that the energy gap scales as Eq.(\ref{e5}) by critical exponent
\begin{eqnarray}
\nu(H)=\frac{2}{4-\alpha(H)},
\label{gafnew}
\end{eqnarray}
which is also obtained with the bosonization technique in Ref.[7]. For example, in the absence of a uniform magnetic field,
$\Delta\sim h^{\frac{2}{3}}$, in agreement with the bosonization and experimental results. Since, by increasing the uniform field $H$,
$\alpha(H)$ decreases, thus we can conclude that the critical gap
exponent $\nu(H)$ drops with increasing uniform field $H$.

To make a numerical check on effect of the uniform field on the
energy gap we have implemented the modified Lanczos
algorithm \cite{grosso} on
 finite-size chains ($N=12, 14, ..., 24$) using periodic
 boundary conditions to calculate the energy gap. We have computed
 the energy gap for different chain lengths in the cases of the uniform fields $0\leq H<0.1$.
 The energy gap as a function of the chain length ($N$), uniform ($H$) and staggered ($h$) fields is defined as
\begin{eqnarray}
\Delta (N, h, H)=E_{1}(N, h, H)-E_{0}(N, h, H),
\end{eqnarray}
where $E_{0}$ is the ground state energy and $E_{1}$ is the first
excited state. In the absence of staggered field ($h=0$) , the
spectrum of the AF Heisenberg model is gapless up to $H=2 J$. The
gap vanishes in the thermodynamic limit proportional to the
inverse of the chain length \cite{fledderjohann2}
\begin{equation}
\lim_{N\rightarrow\infty}\Delta(N, h=0, H){\longrightarrow}
\frac{A(H)}{N}. \label{gaf}
\end{equation}
The coefficient A is known exactly from the Bethe ansatz solution
\cite{klumper} and also can be computed in principle by the
methods of conformal invariance and finite-size scaling
\cite{alcaraz2, woynarovich, alcaraz3}.

When the staggered field is applied, a non zero gap develops. Thus
the staggered field $h_{c}=0$ is a critical point for our model.
In general, the critical point of an infinite system is defined,
in the Hamiltonian formulation, as the value of $h$, $h_{c}$, at
which the mass gap $\Delta(h, H)$ vanishes as Eq.(\ref{e5}). With our Lanczos
scheme we can compute $\Delta(N, h, H)$, which approaches
$\Delta(h, H)$ when N is large. The natural measure of the
deviation of the finite system from the infinite one is
$\frac{L}{L_{0}}$, where $L$ is the linear dimension of the finite
system ($L=N a$, a is the lattice spacing) and $L_{0}$ is the
correlation length of the infinite system ($L_{0}=\xi a$). Thus,
we assume that $\Delta(N, h, H)$ depends on $h$ through
$\frac{L}{L_{0}}$ as
\begin{equation}
\Delta(N, h, H)\sim f(\frac{L}{L_{0}})=f(x),
\end{equation}
where $x=N h^{\nu(H)}$ is a scaling parameter, and $f(x)$ is the scaling function. As expected, this equation
behaves in the combined limit 
\begin{equation}
N\longrightarrow \infty, ~~~~~h\longrightarrow 0~~~~~( x\gg1)
\end{equation}
as Eq.($\ref{e5}$), thus we assume the asymptotic form of the
scaling function $f(x)$
\begin{equation}
f(x) \sim x^{\phi}. \label{phi}
\end{equation}
In addition, we need a factor to cancel the N-dependence of $f(x)$
as $N\longrightarrow\infty$. This factor must be of the form
$N^{-1}$. Thus, we have
\begin{equation}
\Delta(N, h, H)\sim N^{-1} f(x)\label{dgap},
\end{equation}
If we multiply both sides of Eq.(\ref{dgap}) by $N$ we get
\begin{equation}
\lim_{N \to \infty (x \gg 1)} N \Delta(N, h, H) \sim x.
\label{ngap}
\end{equation}
Eq.(\ref{ngap}) shows that the large-$x$ behavior of $N \Delta(N, h,
H)$ is linear in $x$ where the scaling exponent of the energy gap
is $\nu(H)$.
\begin{figure}[tbp]
\centerline{\includegraphics[width=8cm,angle=0]{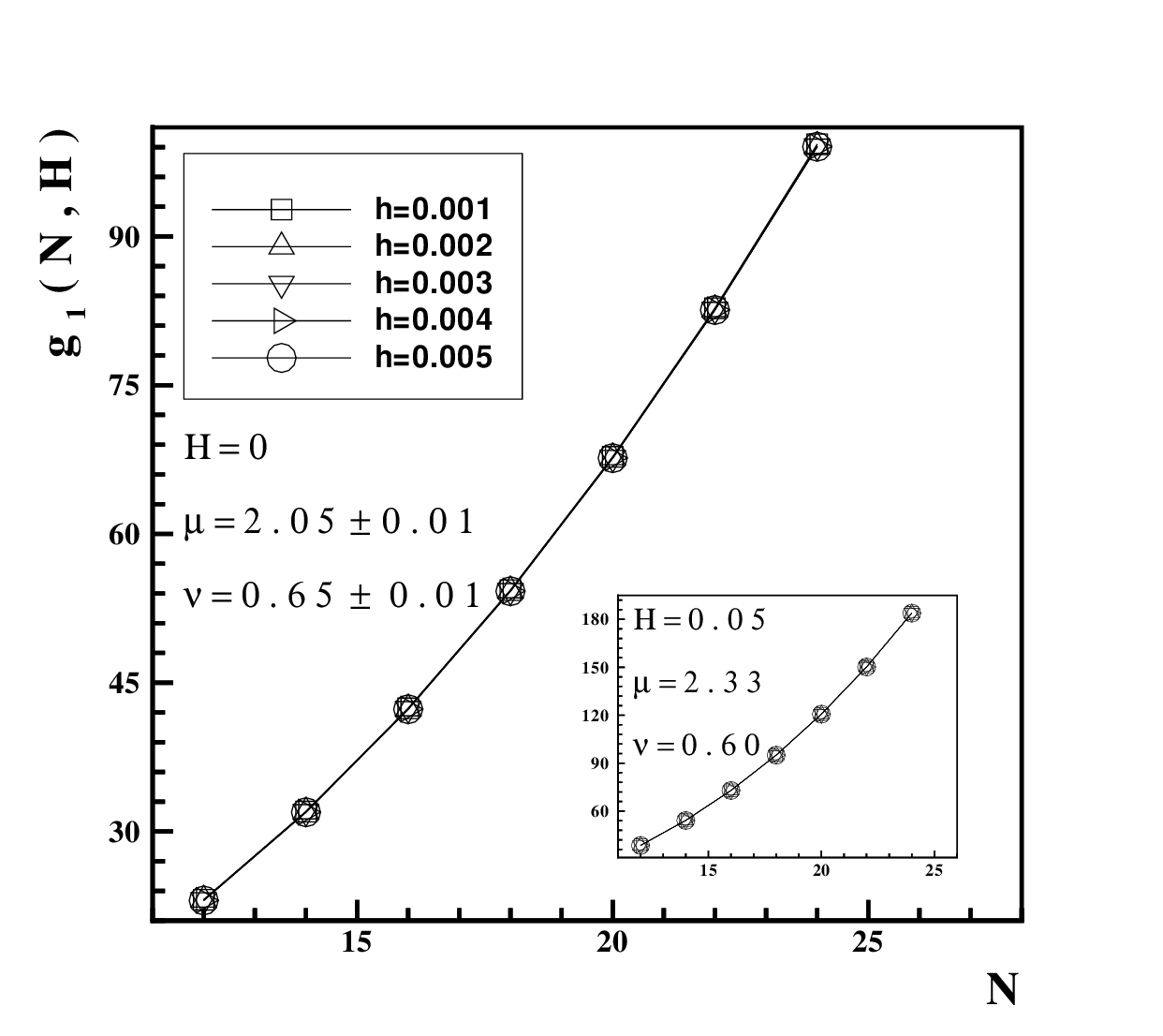}}
\caption{ The value of scaling function $g_{1}(N, H)$ at the fixed
uniform field $H=0$, versus the chain length $N=12, 14, ..., 24$.
The best fit is obtained by $\nu(H=0)=2.05\pm0.01$. In the inset,
the function $g_{1}(N, H)$ is plotted versus N at the uniform
field $H=0.05$ and the best fit is obtained by
$\nu(H=0.05)=2.33\pm0.01$. Data for different staggered fields
$0.001\le h \le 0.005$ fall exactly on each other.} \label{fig1}
\end{figure}
\section{Small-x Regime}
Since the scaling of the gap can only be observed in the
thermodynamic limit and for very small value of $h$, we compute the
energy gap of the model for several values of staggered field
$0.001 \leq h \leq 0.01$ and different chain lengths $N=12, 14,
..., 24$ for fixed uniform fields $0\leq H<0.1$. We have plotted
the values of $N \Delta(N, h, H)$ versus $N h^{\nu(H)}$ for $H=0,
0.03, 0.05, 0.07, 0.09$. The results have been computed on a chain
with the periodic boundary conditions. According to
Eq.(\ref{ngap}), we have found from our numerical results that the
linear behavior is very well satisfied by $\nu(H)\cong2.0$
independent of H. This is very far from the correct value of
critical exponent $\nu(H)\leq\frac{2}{3}$
(Eq.($\ref{gafnew}$)).

Note that the horizontal axes values in the small-$x$ regime are
limited to very small values of $x=N h^{\nu(H)}<0.0024$. Thus, we
are not allowed to obtain the real scaling exponent of the gap
which exists in the thermodynamic limit ($N\longrightarrow \infty$
or $x\gg1$).

When $x$ is small, or in other words $h$ is very small, we might be
away from the thermodynamic behavior to observe the correct
scaling. For very small $h$ in the finite size systems ($N\sim24$)
the value of $x$ will be small ($x\ll1$), which avoid us to get the
information on the large-$x$ behavior of the scaling function
 $f(x)$. In this case, the values of the energy gap coming from a
finite system are basically representing the perturbative behavior
\cite{langari}, which we reproduce for the convenience .

We start from the Hamiltonian Eq.($\ref{efh6}$). The energy
eigenstates of the $\hat{H}_{0}$ carry momentum $p=0$ or $p=\pi$
\begin{eqnarray}
T\mid \psi_{n}(h=0, H)\rangle=\pm \mid \psi_{n}(h=0, H)\rangle,
\end{eqnarray}
where, T is translation operator and $\{\mid \psi_{n}(h=0,
H)\rangle\}$ are eigenstates of unperturbed Hamiltonian
$\hat{H}_{0}$. Since the operator $\sum_{j} (-1)^{j}S_{j}^{z}$
changes the momentum of the state by $\pi$, we obtain that
\begin{eqnarray}
\langle \psi_{n}(0, H) \mid V \mid \psi_{n}(0, H)\rangle=0
\end{eqnarray}
Thus, the gap can be rewritten in the following form
\begin{eqnarray}
\Delta(N, h, H)&=&\Delta(N, 0, H)+g_1(N, H) h^2+   \nonumber \\
&\cdots&+g_n(N, H) h^{2n}, \label{pertexp}
\end{eqnarray}
where $n$ is an integer. It is a good approximate to neglect the
effect of higher order terms for $h\leq 0.01$. Because the second-order perturbation correction is not zero in
the staggered field, the leading nonzero term is $h^{2}$. If the
small-$x$ behavior of the scaling function is defined as $f(x)\sim
x^{\phi_{s}}$, we find that
\begin{eqnarray}
\nu(H) \phi_{s}=2 \label{nu}.
\end{eqnarray}
This shows that in the small-$x$ regime, $N\Delta(N, h, H)$ is a
linear function of $x^{\frac{2}{\nu(H)}}$. This is in agreement
with our data in small-$x$ regime, where $\phi_{s}=1$ and according
to Eq.($\ref{nu}$), the value of $\nu(H)$ is found to be
$\nu(H)=2.0$.

Let us consider the large-$N$ behavior of $g_{1}(N, H)$ at fixed $H$ as
\begin{eqnarray}
\lim_{N\rightarrow\infty} g_1(N, H) \simeq a_1(H) N^{\mu(H)}.
\label{e2}
\end{eqnarray}
This leads to
\begin{eqnarray}
\Delta(N, h, H)\simeq \frac{A(H)}{N}(1+b_{1}(H) N^{\mu(H)+1}
h^{2}),\label{e24}
\end{eqnarray}
where $b_{1}(H)$ is a constant (at fixed $H$). We can write Eq.
($\ref{e24}$) in terms of the scaling variable $x$
\begin{eqnarray}
N \Delta(N, h, H)\simeq A(H)(1+N^{\mu(H)+1-\frac{2}{\nu(H)}}
x^{\frac{2}{\nu(H)}})
\end{eqnarray}
For large-$N$ limit this equation should be independent of $N$,
leading to the relation between $\mu(H)$ and $\nu(H)$ as
\begin{eqnarray}
\nu(H)=\frac{2}{\mu(H)+1}\label{s26}
\end{eqnarray}
The above arguments propose to look for the large-$N$ behavior of
$g_{1}(N, H)$. To determine the $\mu(H)$ exponent, we have plotted
in Fig.1 the following expression versus $N$
\begin{eqnarray}
g_{1}(N, H)\simeq\frac{\Delta(N, h, H)-\Delta(N, 0, H)}{h^{2}}
\end{eqnarray}
for fixed values of staggered field $h$ ($0.001\leq h \leq 0.005$),
and different sizes, $N=12, 14, ..., 24$ at the uniform field
$H=0$. We found the best fit to our data for
$\mu(H=0)=2.04\pm0.01$. The inset in Fig.1 shows the $g_{1}(N,
H)$ versus $N$ at fixed $H=0.05$. In this case, the best fit, found
for $\mu(H=0.5)=2.33\pm0.01$. Our data for different $h$ values,
fall perfectly on each other, which shows that our results for
$g_{1}(N, H)$ in fixed uniform field $H$, are independent of the
staggered field $h$ as expected. By using Eq.($\ref{s26}$) we have
found, $\nu(H=0)=0.66\pm0.01$ and $\nu(H=0.05)=0.60\pm0.01$. We
have also implemented our numerical tool to calculate the critical exponent $\nu(H)$ at $H=0.03, 0.07, 0.09$. The results have been
presented in Table I.
\begin{figure}[tbp]
\centerline{\includegraphics[width=8cm,angle=0]{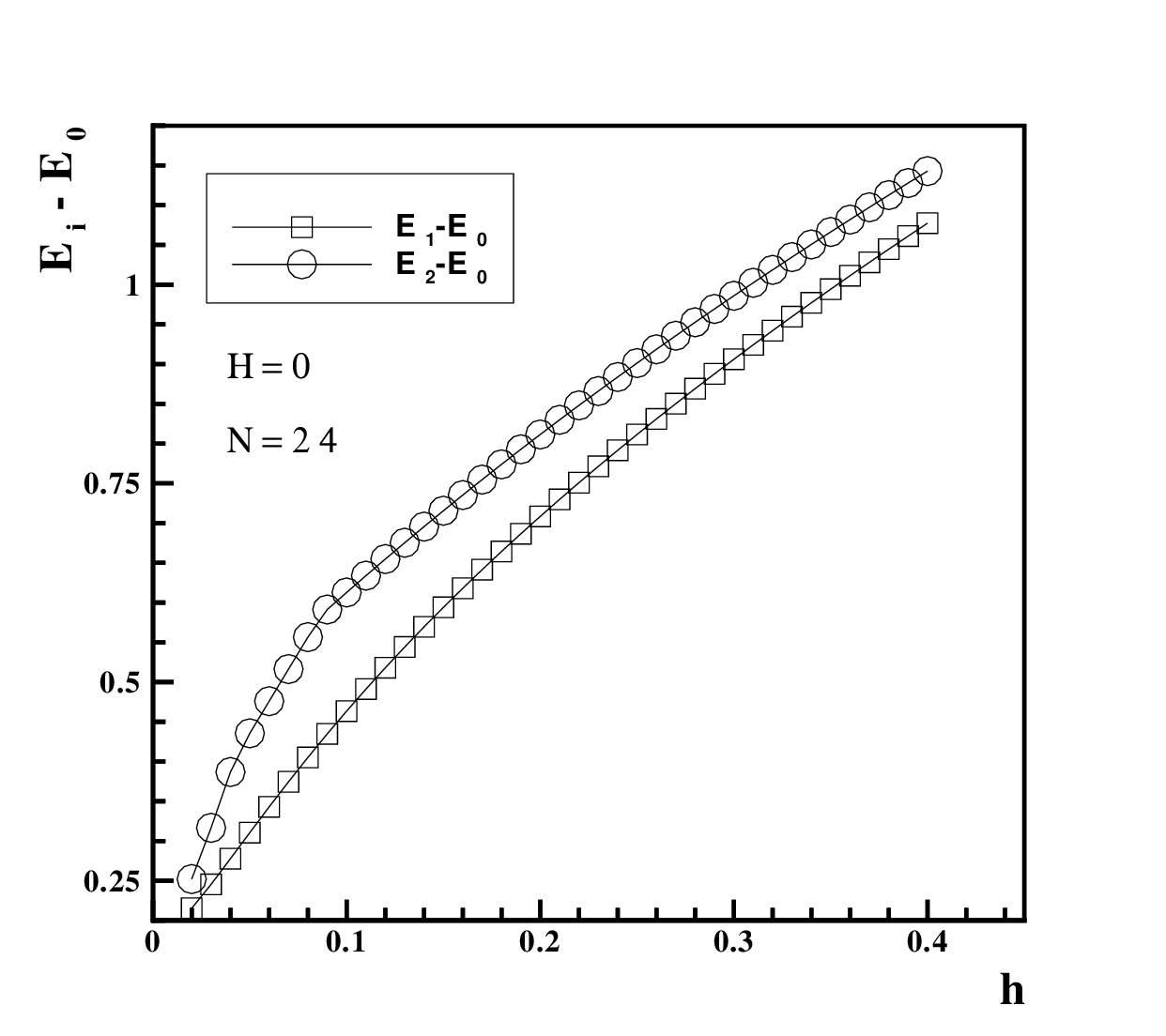}}
\caption{ Difference between the two lowest energy levels and the
ground state energy as a function of the staggered magnetic field
for finite chain length $N=24$ and $H=0$ in the region $0.01\leq h
\leq 0.4$.} \label{fig2}
\end{figure}
\begin{figure}[tbp]
\centerline{\includegraphics[width=8cm,angle=0]{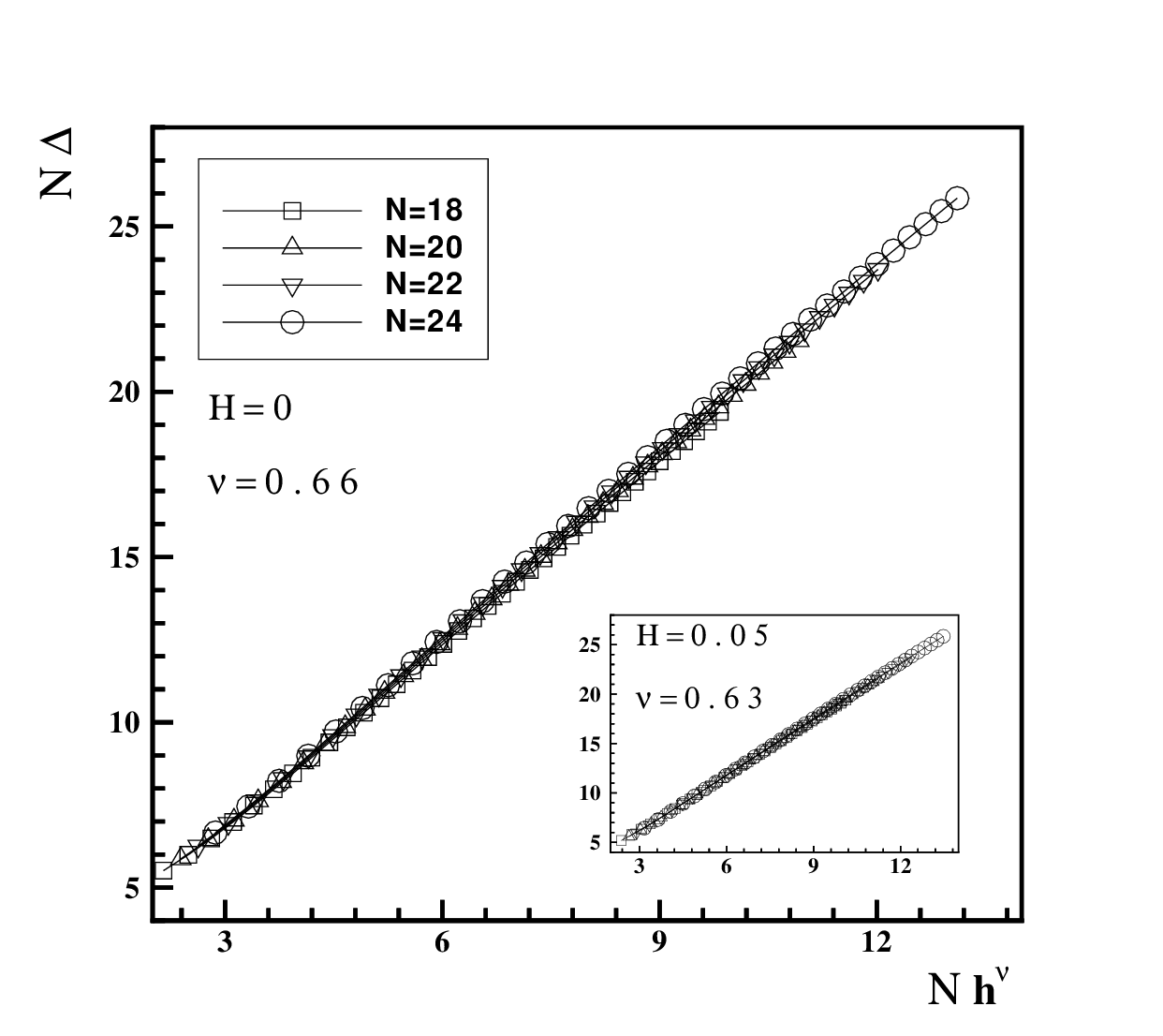}}
\caption{ The product of energy gap and chain length ($N \Delta$)
versus $N h^{\nu(H)}$ at the uniform field $H=0$. In the range of
staggered field $0.04 \leq h \leq 0.4$, linear behavior is
obtained by choosing $\nu(H=0)=0.66$ for all different chain
lengths $N=18, 20, 22, 24$. In the inset, $N \Delta$ is plotted
for uniform field $H=0.05$. The linear behavior is obtained by
$\nu(H=0.05)=0.63$. Data for different chain lengths fall on each
other.} \label{fig3}
\end{figure}
\begin{figure}[tbp]
\centerline{\includegraphics[width=8cm,angle=0]{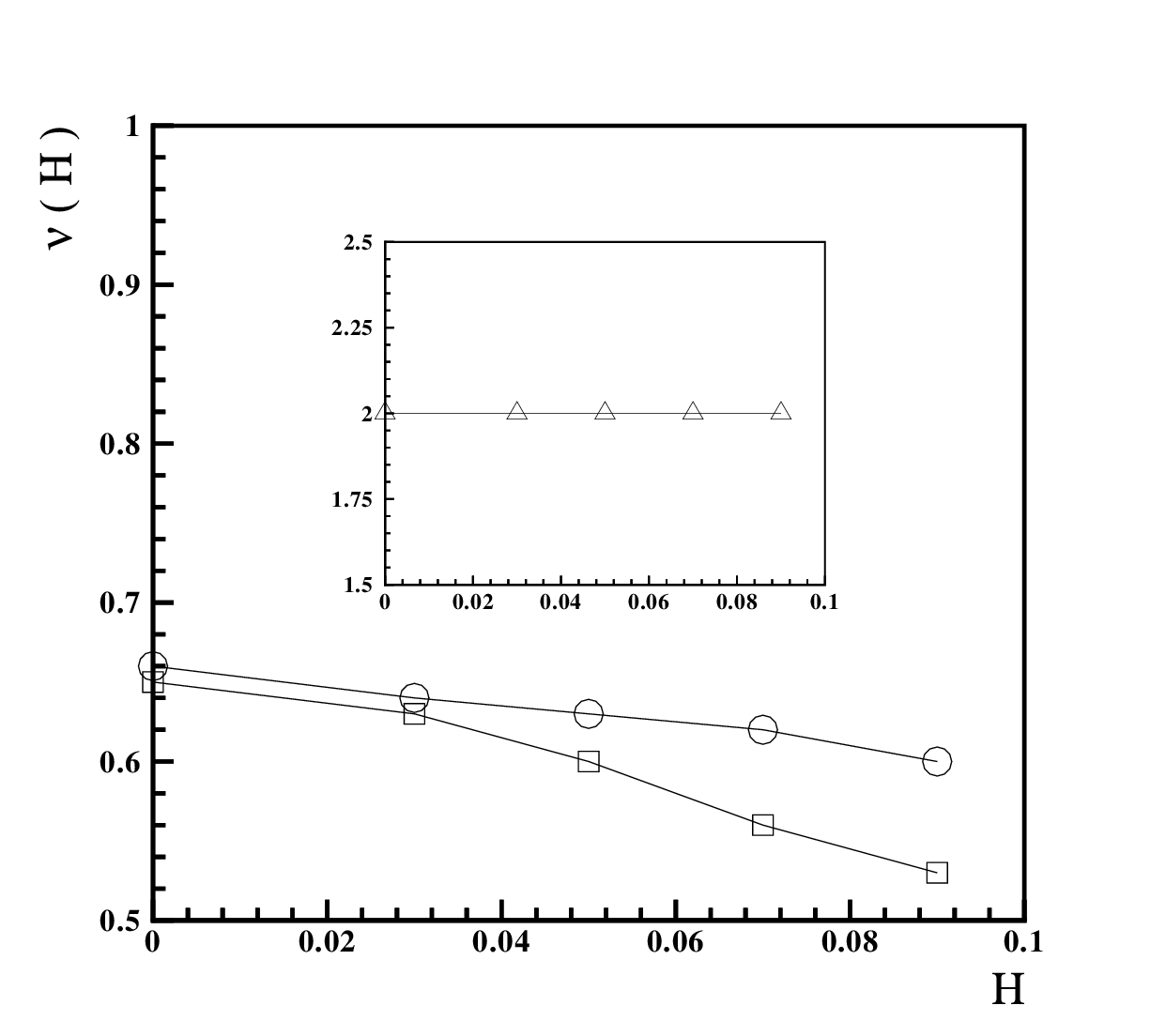}}
\caption{ A graph of the critical exponent $\nu$ versus uniform field $H$. Both critical exponents $\nu_{s}$ (squares) and $\nu_{l}$ (circles) start at the known value $\frac{2}{3}$ and then drop with the increasing uniform field $H$. In the inset, we have plotted the critical exponent $\nu$ which is obtained directly by extrapolating the numerical results of the energy gap in the range of the staggered field 
$0.001\leq h \leq 0.01$. } \label{fig4}
\end{figure}
\begin{table}[t]
\begin{center}
\label{table2} \caption{The exponent of $g_{1}(N, H)$ and gap
exponent versus different values of the uniform field H in small-x
and large-x regimes.  }
\begin{tabular}
{cccc} \hline \hline
$H$ ~~~~~~~~~~~~~~~~~ & $\mu$~~~~~~~~~~~~~~~~~ & $\nu_{S}$~~~~~~~~~~~~~~~~~ &  $\nu_{L}$\\
\hline
$0.0$ ~~~~~~~~~~~~~~~~~   & $2.05$~~~~~~~~~~~~~~~~~ &$0.65$~~~~~~~~~~~~~~~~~   &$0.66$  \\
$0.03$ ~~~~~~~~~~~~~~~~~ &$2.20$~~~~~~~~~~~~~~~~~ &$0.63$~~~~~~~~~~~~~~~~~   &$0.64$  \\
$0.05$ ~~~~~~~~~~~~~~~~~ &$2.33$~~~~~~~~~~~~~~~~~ &$0.60$~~~~~~~~~~~~~~~~~   &$0.63$  \\
$0.07$ ~~~~~~~~~~~~~~~~~ &$2.58$~~~~~~~~~~~~~~~~~ &$0.56$~~~~~~~~~~~~~~~~~   &$0.62$  \\
$0.09$ ~~~~~~~~~~~~~~~~~ &$2.80$~~~~~~~~~~~~~~~~~ &$0.53$~~~~~~~~~~~~~~~~~   &$0.60$  \\
\hline \hline
\end{tabular}
\end{center}
\end{table}



\section{large-x regime}

In our numerical calculations because of memory issues, we were
limited to consider the maximum chain length $N=24$. Therefore
the value of $x$ cannot be increased by increasing the size of
chain. The problem appears if the calculation is done by density
matrix renormalization group (DMRG) method. In that case, we may
extend the calculation to larger sizes, $N\sim200$, which cannot
increase $x$ much larger than one (for $0.001<h<0.01$). On the other hand
we may increase the staggered field for increasing $x$. But we
should note that, in general there is usually level crossing between the energy levels in
finite size systems. Which can change the behavior of the gap
\cite{dmitriev} and lead to incommensurate effects. As an example,
the dependences on the excitation energies of the three lowest
levels on the staggered field $h$ ($0.01\leq h \leq0.4$) are shown
in Fig.2 for $N=24$ and $H=0$. From this figure, it can be seen
that the two lowest excited states do not cross each other in this case. This
means that we can increase the value of $x$ by increasing $h$ up to
$h=0.4$. Since that the regime where we can observe the right
scaling of the gap is in very small value of $h$, we have restricted
our numerical computations to upper limit of staggered field
$h=0.4$. In this respect, we have performed some numerical
computations on the Hamiltonian Eq.($\ref{Hamiltonian}$) for
large-$x$ values, and the results are plotted in Fig.3. In this
figure the range of the staggered field is $0.04\leq h \leq 0.4$,
which causes that we get large-$x$ values for the considered chain
lengths $N=18, 20, 22, 24$ and $ H=0$. The inset in Fig.3 shows
$N\Delta(N, h, H)$ versus $x=N h^{\nu(H)}$ at fixed uniform field
$H=0.05$. In this case, we have obtained, $\nu(H=0)=0.66$ and
$\nu(H=0.05)=0.63$. It should be mentioned that if we choose
another value for $\nu$ the plot will not be linear in $x$. Also,
our data for different size chains fall perfectly on each other,
which is expected from the scaling behavior.

We have extended our numerical computations to consider values of
the uniform field in the small region $0\leq H <0.1$. The results
have been presented in Table I. We have listed $\mu$, the
resulting $\nu_{s}$ that is obtained from perturbative approach,
and result of the large-$x$ regime $\nu_{l}$ for different values
of the uniform field $H$. In Fig.4 we have plotted the critical exponent $\nu(H)$ versus the uniform field $H$. As it is clearly seen from this figure, $\nu_{s}$ and $\nu_{l}$ start at the known value $\frac{2}{3}$ and then drop with the increasing uniform field $H$. The exponents are in good agreement with each other and show good
agreement with the exponents derived in the field theoretical
approach (Eq.($\ref{gafnew}$)). The inset in Fig.4 shows the H-dependence of the critical exponent $\nu$ which is obtained directly by extrapolating the numerical results of the energy gap in the range of the staggered field 
$0.001\leq h \leq 0.01$ . It is clearly seen from this figure that the behavior of the gap in finite systems is different from its behavior at the thermodynamic limit, and with respect to the range of the staggered magnetic field the behavior of the gap deviates from the predicted scaling behavior. 

On the other hand, Fouet et.al studied \cite{fouet} the gap-induced by the staggered field $h$ at the saturation uniform field $H_{c}=2 J$. Using field theoretical arguments, they found that the gap scales as $\Delta(h, H_{c})\sim h^{4/5}$. By applying the DMRG method for system sizes up to $N=100$, they have also computed the exponent of the energy gap $0.81$. We have extended our numerical computations to consider the uniform field at the saturation value $H_{c}$. In this case, we have obtained, $\nu_{s}(H_{c})=0.78$ and $\nu_{l}(H_{c})=0.82$ in good agreement with the Foet results.

\section{conclusions}

In this paper, we have studied the energy gap of the 1D
AF-Heisenberg model in the presence of both uniform ($H$) and
staggered ($h$) magnetic fields using the exact diagonalization
technique. We have implemented the modified Lanczos method to
obtain the excited state energies with the same accuracy as the
ground state one. We have been limited to a maximum of $N = 24$, because of memory
considerations. We have shown, if the energy gap in the thermodynamic limit had been obtained by extrapolating the numerical results for finite systems, then the behavior of the gap may deviate from the predicted scaling behavior. This deviation depends on the range of the staggered magnetic field ($h$). We have found in the range of very small values of the staggered magnetic field ($0.001\leq h \leq 0.01$), the values of the energy gap coming from a finite system basically represent the perturbative behavior. We have shown that in this range of the staggered magnetic field, we are not
allowed to read the scaling exponent of the energy gap directly from the extrapolated numerical results. 

We have applied a general finite size scaling procedure for
investigating the $H$-dependence of the critical exponent of the gap. We have identified two
regimes through which the real critical exponent $\nu(H)$ can be
numerically calculated. To find the correct exponent of the gap in
small-$x$ regime ($x=N h^{\nu(H)}\ll1$), we have used the scaling
behavior of the coefficient of the leading term in the
perturbation expansion, which is introduced by authors in
Ref.[25]. In the large-$x$ regime using the standard finite size
scaling Eq.(\ref{ngap}), we have computed the correct critical exponent. On the other hand, using the conformal estimates of the
small perturbation ($h\ll1$), we have found the $H$-dependence of
the critical exponent (Eq.(\ref{gafnew})) from the theoretical
point of view. Our numerical results in both regimes are in well
agreement with the results obtained by the theoretical and
numerical approaches.
\section{Acknowledgments}
I would like to thank A. Langari, M. R. H. Khajehpour, G. I. Japaridze, J. Abouie and M. Kohandel for insightful comments and fruitful discussions that led to an improvement of this work. 

\medskip



\begin{thebibliography}{99}


\bibitem{dender}
D. C. Dender, P. R. Hammar, D. H. Reich, C. Broholm and G. Aeppli,
Phys. Rev. Lett. {\bf 79}, 1750 (1997).
\bibitem{asano1}
T. Asano, H. Nojiri, Y. Inagaki, J. P. Boucher, T. Sakon, Y. Ajiro
and M. Motokawa, Phys. Rev. Lett. {\bf 84}, 5880 (2000).
\bibitem{asano2}
T. Asano, H. Nojiri, W. Hiegemoto, A. Koda, R. Kadono and Y.
Ajiro, J. Phys. Soc. Jpn. {\bf 72}, 594 (2002).
\bibitem{oshikawa1}
M. Oshikawa, K. Ueda, H. Aoki, A. Ochiai and M. Kohgi, J. Phys.
Soc. Jpn. {\bf 68}, 3181 (1999).
\bibitem{shiba}
H. Shiba, K. Ueda and O. Sakai, J. Phys. Soc. Jpn. {\bf 69}, 1493
(2000).
\bibitem{kohgi}
M. Kohgi, K. Iwasa, J. M. Mignot, B. Fak, P. Gegenwart, M. Lang,
A. Ochiai, H. Aoki and T. Suzuki, Phys. Rev. Lett. {\bf 86}, 2439
(2001).
\bibitem{oshikawa2}
M. Oshikawa and I. Affleck, Phys. Rev. Lett. {\bf 79}, 2883
(1997); Phys. Rev. {\bf B 60}, 1038 (1999).
\bibitem{wang}
Y. -J. Wang, F. H. L. Essler, M. Fabrizio and A. A. Nersesyan,
Phys. Rev. {\bf B 66}, 024412 (2002).
\bibitem{sato}
M. Sato and M. Oshikawa, Phys. Rev. {\bf B 69}, 054406 (2004)
\bibitem{alcaraz1}
F. C. Alcaraz and A. L. Malvazzi, J. Phys. {\bf A 28}, 1521
(1995).
\bibitem{feyerherm}
R. Feyerherm, S. Abens, D. Gunther, T. Ishida, M. Meibner, M.
Meschke, T. Nogami and M. Steiner, J. Phys. Condens. Matter {\bf
12}, 8495 (200).
\bibitem{pokrovsky}
V. L. Pokrovsky and A. L. Talapov, Soc. Phys. JETP {\bf 51}, 134
(1980).
\bibitem{griffiths}
R. B. Griffiths, Phys. Rev. {\bf 133}, A768 (1964).
\bibitem{lou}
J. Lou, C. Chen, J. Zhao, X. Wang, T. Xiang, Z. Su and L. Yu,
Phys. Rev. Lett. {\bf 94}, 217207 (2005).
\bibitem{fledderjohann1}
A. Fledderjohann, M. Karbach and K.-H. Mutter, Eur. Phys. J. {\bf
B 7}, 225 (1999).
\bibitem{chernyshev}
A. L. Chernyshev, Phys. Rev. {\bf B 72}, 174414 (2005).
\bibitem{bogoliubovn}
N. M. Bogoliubov, A. G. Izergin and V. E. Korepin, Nucl. Phys.
{\bf B 275}, 687 (1986).
\bibitem{grosso}
G. Grosso and L. Martinelli, Phys. Rev. {\bf B 51}, 13033 (1995).
\bibitem{fledderjohann2}
A. Fledderjohann, M. Karbach and K.-H. Mutter, Eur. Phys. J. {\bf
B 5}, 487 (1998).
\bibitem{klumper}
A. Klumper, Eur. Phys. J. {\bf B 5}, 677 (1998), and references
therein.
\bibitem{alcaraz2}
F. C. Alcaraz, M. N. Barber, M. T. Batchelor,  Phys. Rev. Lett.
{\bf 58}, 771 (1987).
\bibitem{woynarovich}
F. Woynarovich, H. -P. Eckle, J. Phys. {\bf A }: Math. Gen. 20,
{\bf L97} (1987).
\bibitem{alcaraz3}
F. C. Alcaraz, M. N. Barber, M. T. Batchelor, Ann. Phys. {\bf
182}, 280 (1988).

\bibitem{dmitriev}
D. V. Dmitriev, V. Ya. Krivnov, A. A. Ovchinikov, and A. Langari,
JETP {\bf 95}, 538 (2002).
\bibitem{langari} A. Langari and S. Mahdavifar, Phys. Rev. {\bf B 73},
54410 (2006).
\bibitem{fouet} J. -B. Fouet, O. Tchernyshyov, and F. Mila, Phys. Rev. {\bf B 70},
174427 (2004).












\end{thebibliography}
\end{document}